\pdfoutput=1

\documentclass[11pt]{article}

\usepackage{acl}

\usepackage{times}
\usepackage{latexsym}

\usepackage[T1]{fontenc}

\usepackage[utf8]{inputenc}

\usepackage{microtype}
\usepackage{amsmath}
\usepackage{graphicx}
\usepackage{booktabs}
\usepackage{multirow}
\usepackage{xspace}
\usepackage{mathrsfs}

\usepackage{tikz}
\usepackage{xcolor}
\newcommand*\circled[1]{\tikz[baseline=(char.base)]{
            \node[shape=circle,fill,inner sep=0.2pt] (char) {\textcolor{white}{#1}};}}

\newcommand{\tool}{CoSHC\xspace} 
\newcommand{\Sec}{Sec.\xspace}

\usepackage{color}

\newcommand{\saveSpaceFig}{\vspace{-3pt}}
\newcommand{\saveSpaceText}{\vspace{-1pt}}

\usepackage{enumitem}
\usepackage{arydshln}
\title{Accelerating Code Search with Deep Hashing and Code Classification}

\newcommand\corrauthorfootnote[1]{%
  \begingroup
  \renewcommand\thefootnote{}\footnote{\textsuperscript{\dag}#1}%
  \addtocounter{footnote}{-1}%
  \endgroup
}

\author{Wenchao Gu$^1$\thanks{\ \ \ Work done while this author was an intern at Microsoft Research. Wenchao Gu (wcgu@cse.cuhk.edu.hk).}, Yanlin Wang$^{2,\dag}$, Lun Du$^2$, Hongyu Zhang$^3$, \\ \textbf{ Shi Han$^2$, Dongmei Zhang$^2$, and Michael R. Lyu$^1$}\\ 
    $^1$ Department of Computer Science and Engineering, \\ The Chinese University of Hong Kong, China.\\	
	$^2$ Microsoft Research Asia, Beijing, China\\
	$^3$ The University of Newcastle, Australia \\
}

\begin{document}
\maketitle
\begin{abstract}
Code search is to search reusable code snippets from source code corpus based on natural languages queries. 
Deep learning-based methods on code search have shown promising results. However, previous methods focus on retrieval accuracy, but lacked attention to the efficiency of the retrieval process. 
We propose a novel method CoSHC to accelerate code search with deep hashing and code classification, aiming to perform efficient code search without sacrificing too much accuracy. 
To evaluate the effectiveness of CoSHC, we apply our method on five code search models. Extensive experimental results indicate that compared with previous code search baselines, CoSHC can save more than 90\% of retrieval time meanwhile preserving at least 99\% of retrieval accuracy.
\end{abstract}

\section{Introduction}
\label{sec:intro}
Code reuse\corrauthorfootnote{ Yanlin Wang is the corresponding author (yanlwang@microsoft.com).} is a common practice during software development process. It improves programming productivity as developers' time and energy can be saved by reusing existing code. According to previous studies~\citep{BrandtGLDK09,LvZLWZZ15}, many developers tend to use natural language to describe the functionality of desired code snippets and search the Internet/code corpus for code reuse.

Many code search approaches~\citep{BrandtGLDK09, McMillanGPXF11, LvZLWZZ15, du2021single} have been proposed over the years. With the rapid growth of open source code bases and the development of deep learning technology, recently deep learning based approaches have become popular for tackling the code search problem~\citep{GuZ018, abs-1909-09436, GuLGWZXL21}. 
Some of these approaches adopt neural network models to encode source code and  query descriptions into representation vectors in the same embedding space. The distance between the representation vectors whose original code or description are semantically similar should be small. Other approaches~\citep{FengGTDFGS0LJZ20,GuoRLFT0ZDSFTDC21,du2021single} regard the code search task as a binary classification task, and calculate the probability of code matching the query.

In the past, deep learning-based methods focused on retrieval accuracy, but lacked attention to the efficiency of retrieval on large-scale code corpus. However, both types of these deep learning-based approaches directly rank all the source code snippets in the corpus during searching, which will incur a large amount of computational cost. For the approaches that separately encode code and description representation vectors, the similarity of the target query vector with all code representation vectors in the corpus needs to be calculated for every single retrieval. In order to pursue high retrieval accuracy, a high dimension is often set for the representation vectors. For example, in CodeBERT, the dimension of the final representation vector is 768. The similarity calculation between a pair of code and query vectors will take 768 multiplications and 768 additions between two variables with double data type. The total calculation of single linear scan for the whole code corpus containing around 1 million code snippets is extremely large - around 1 billion times of multiplications and additions. As for the approaches adopting binary classification, there is no representation vectors stored in advance and the inference of the target token sequence with all the description token sequences needs to be done in real time for every single retrieval. Due to the large number of parameters in the current deep learning models, the computation cost will be significant. 

Hashing is a promising approach to improve the retrieval efficiency and widely adopted in other retrieval tasks such as image-text search and image-image search. Hashing techniques can convert high dimensional vectors into low dimensional binary hash code, which greatly reduce the cost of storage and calculation~\citep{abs-2003-03369}. 
Hamming distance between two binary hash code can also be calculated in a very efficient way by running XOR instruction on the modern computer architectures~\citep{WangLKC16}.
However, the performance degradation is still not avoidable during the conversion from representation vectors to binary hash codes even the state-of-the-art hashing models are adopted. The tolerance of performance degradation from most users is quite low and many of them are willing to sweep the performance with efficiency. In order to preserve the performance of the original code search models that adopt bi-encoders for the code-query encoding as much as possible, we integrate deep hashing techniques with code classification, which could mitigate the performance degradation of hashing model in the recall stage by filtering out the irrelevant data.

Specifically, in this paper, we propose a novel approach \textbf{\tool} (Accelerating Semantic \underline{Co}de \underline{S}earch with Deep \underline{H}ashing and Code \underline{C}lassification) for accelerating the retrieval efficiency of deep learning-based code search approaches. \tool firstly clusters the representation vectors into different categories. It then generates binary hash codes for both source code and queries according to the representation vectors from the original models. Finally, \tool gives the normalized prediction probability of each category for the given query, and then \tool will decide the number of code candidates for the given query in each category according to the probability. Comprehensive experiments have been conducted to validate the performance of the proposed approach. {The evaluation results show that \tool can preserve more than 99\% performance of most baseline models.} 
We summarize the main contributions of this paper as follows:
\vspace{-5pt}
\begin{itemize}[leftmargin=10pt]
\setlength{\itemsep}{0pt}
\setlength{\parsep}{0pt}
\setlength{\parskip}{0pt}
\item We propose a novel approach, \tool, to improve the retrieval efficiency of previous deep learning based approaches. \tool is the first approach that adopts the recall and re-rank mechanism with the integration of code clustering and deep hashing to improve the retrieval efficiency of deep learning based code search models.
\item We conduct comprehensive experimental evaluation on public benchmarks. The results demonstrate that \tool can greatly improve the retrieval efficiency meanwhile preserve almost the same performance as the baseline models.
\end{itemize}

\section{Background}
\label{sec:background}

\saveSpaceText
\subsection{Code Search}
\saveSpaceText

In this subsection, we briefly review some deep learning based code search approaches. \citet{SachdevLLKS018} firstly propose the neural network based model NCS to retrieve the source code from a large source code corpus according to the given natural language descriptions. 
\citet{CambroneroLKS019} propose a neural network model UNIF based on bag-of-words, which embeds code snippets and natural language descriptions into a shared embedding space. 
\citet{GuZ018} propose to encode source code representation with API sequences, method name tokens and code tokens. 
\citet{YaoPS19} treat code annotation and code search as dual tasks and utilize the generated code annotations to improve code search performance. 
\citet{abs-1909-09436} explore different neural architectures for source code representation and discover that the self-attention model achieves the best performance. 
\citet{GuLGWZXL21} extract the program dependency graph from the source code and adopt long short term memory (LSTM) networks to model this relationship. 
\citet{FengGTDFGS0LJZ20} propose a pre-trained model for source code representation and demonstrate its effectiveness on the code search task.

\saveSpaceText
\subsection{Deep Hashing}
\saveSpaceText

In this subsection, we briefly introduce some representative
unsupervised cross-modal hashing methods. 
In order to learn a unified hash code, \citet{DingGZ14} propose to adopt collective matrix factorization with latent factor model from different modalities to merge multiple view information sources.
\citet{ZhouDG14} firstly utilize sparse coding and matrix factorization to extract the latent features for images and texts, respectively. Then the learned latent semantic features are mapped to a shared space and quantized to the binary hash codes. \citet{WangOYZZ14} suggest using stacked auto-encoders to capture the intra- and inter-modal semantic relationships of data from heterogeneous sources. 
\citet{0001XLYSS17} and \citet{ZhangPY18} adopt adversarial learning for cross-modal hash codes generation. 
\citet{WuLHLDZS18} propose an approach named UDCMH that integrates deep learning and matrix factorization with binary latent factor models to generate binary hash codes for multimodal data retrieval. 
By incorporating Laplacian constraints into the objective function, UDCMH preserve not only the nearest neighbors but also the farthest neighbors of data. 
Unlike using Laplacian constraints in the loss function, \citet{SuZZ19} construct a joint-semantic affinity matrix that integrates the original neighborhood information from different modalities to guide the learning of unified binary hash codes.

\begin{figure*}[ht]
\centering
\includegraphics[width=0.82\textwidth]{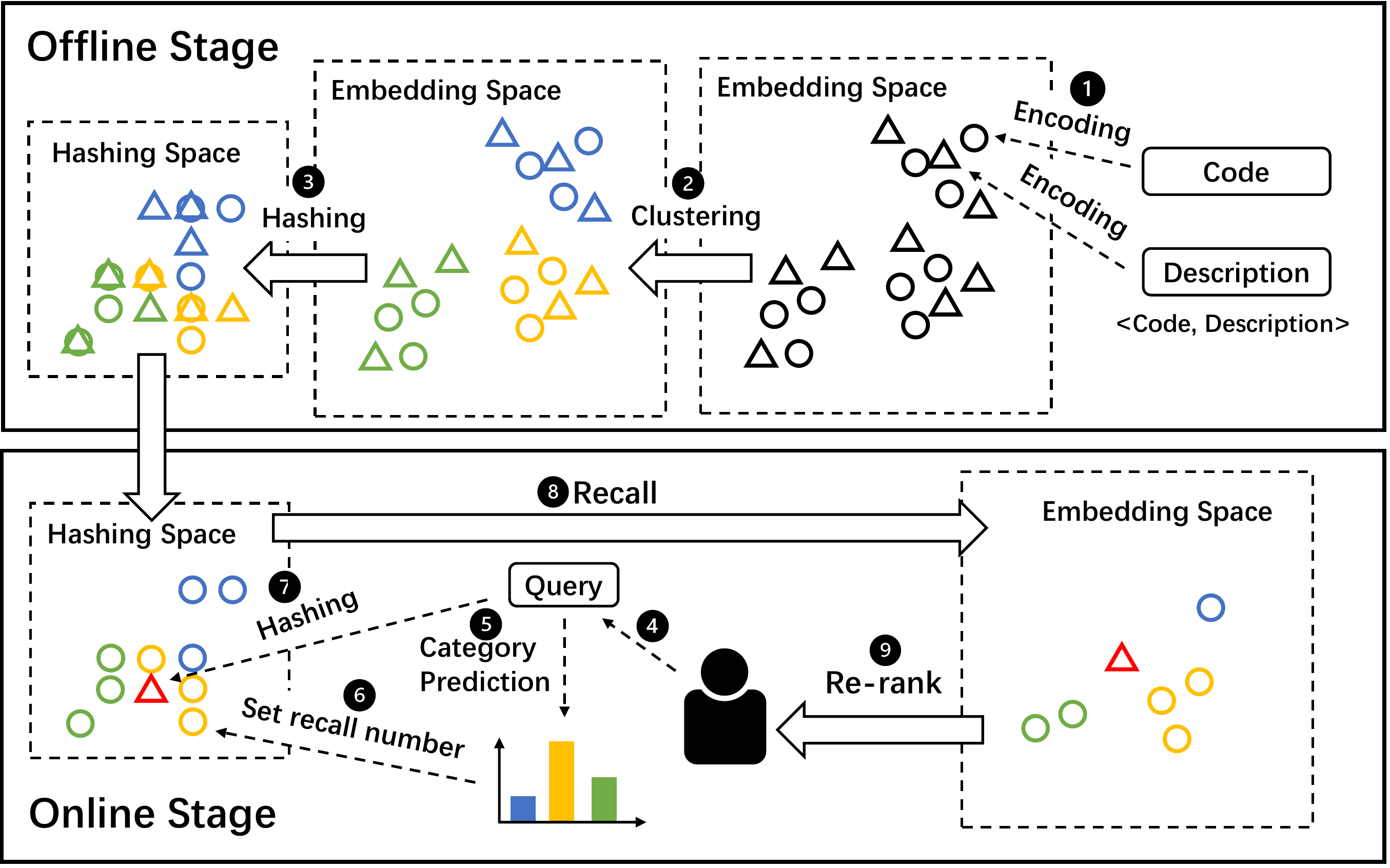}
\saveSpaceFig
\caption{Overview of the proposed \tool. \circled{1} Encoding the code token sequence and description token sequence via original code retrieval models. \circled{2} Clustering the code representation vectors into several categories. \circled{3} Converting the original code representation vectors into binary hash codes.\circled{5}\circled{6} Predicting the category of the query given by users and set the number of code candidates for different categories. \circled{7} Converting the input query into binary hash code. \circled{8} Recall the code candidates according to the hamming distance and the number of code candidates for each category. \circled{9} Re-ranking all the code candidates according to the cosine similarity between the input query description vectors and code candidates' representation vectors and return the results to the user. }
\saveSpaceFig
\label{fig:framework}
\end{figure*}

\saveSpaceText
\section{Method}
\label{sec:method}
\saveSpaceText

We propose a general framework to accelerate existing Deep Code Search (DCS) models by decoupling the search procedure into a recall stage and a re-rank stage. Our main technical contribution lies in the recall stage. Figure~\ref{fig:framework} illustrates the overall framework of the proposed approach. \tool consists of two components, i.e., Offline and Online. In  Offline, we take the code and description embeddings learned in the given DCS model as input, and learn the corresponding hash codes by preserving the relations between the code and description embeddings. In Online, we recall a candidate set of code snippets according to the Hamming distance  between the query and code, and then we use the original DCS model to re-rank the candidates.

\saveSpaceText
\subsection{Offline Stage}
\saveSpaceText

\noindent{\textbf{Multiple Code Hashing Design with Code Classification Module}}
Since the capacity of binary hashing space is very limited compared to Euclidean space, the Hamming distance between similar code snippets will be too small to be distinguishable if we adopt a single Hashing model. To be specific, we cluster the codebase using K-Means algorithm with the code embeddings learned from the given DCS model. The source code whose representation vectors are close to each other will be classified into the same category after the clustering.%

\saveSpaceText

\begin{figure*}[ht]
\centering
\includegraphics[width=1\textwidth]{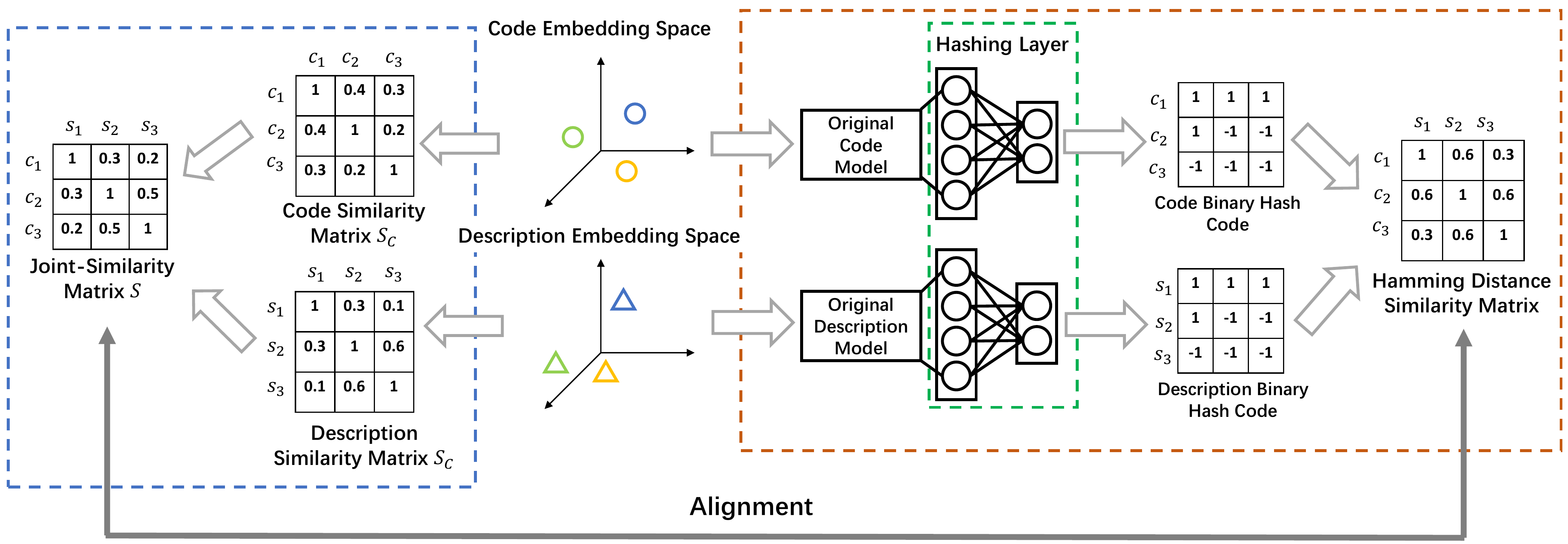}
\saveSpaceFig
\caption{Architecture of the hashing module. The original representation vectors will be utilized for the joint-similarity matrix construction at first. Then the joint-similarity matrix will be utilized as the labels for  training binary hash codes generation. The training objective is to make the Hamming distance similarity matrix to be identical as the joint-similarity matrix.}
\saveSpaceFig
\label{fig:hashing_module}
\end{figure*}

\noindent{\textbf{Deep Hashing Module}}
The deep hashing module aims at generating the corresponding binary hash codes for the embeddings of code and description from the original DCS model. Figure~\ref{fig:hashing_module} illustrates the framework of the deep hashing module. To be specific, three fully-connected (FC) layers with $\rm tanh(\cdot)$ activation function are adopted to replace the output layer in the original DCS model to convert the original representation vectors into a soft binary hash code. 

The objective of the deep hashing module is to force the Hamming distance between hashing representations of code pairs and description pairs approaching the Euclidean distance between the corresponding embeddings. Thus, we need to calculate the ground truth similarity matrix between code pairs and description pairs firstly. For performance consideration, we calculate the similarity matrix within a mini-batch.  

\saveSpaceText

To construct such a matrix, we first define the code representation vectors and the description representation vectors in the original code search model as $V_C=\{v_{c}^{(1)},...,v_{c}^{(n)}\}$ and $V_D=\{v_{d}^{(1)},...,v_{d}^{(n)}\}$ 
, respectively. $V_C$ and $V_D$ represent the representation vectors matrix for the entire batch, while $v_{c}^{(i)}$ and $v_{d}^{(i)}$ represent the representation vector for the single code snippet or query. After normalizing $V_C$, $V_D$ to $\hat{V}_C$, $\hat{V}_D$ with $l_2$-norm, we can calculate the code similarity matrices ${S_C}=\hat{V}_C{\hat{V}_C}^T$ 
and summary similarity matrices ${S_D}=\hat{V}_D{\hat{V}_D}^T$ to describe the similarity among code representation vectors and summary representation vectors, respectively. In order to integrate the similarity information in both $S_C$ and $S_D$, we combine them with a weighted sum:
\begin{equation}
    \tilde{S} = \beta S_C + (1 - \beta)S_D, \beta \in [0,1]
\end{equation}
\noindent where $\beta$ is the weight parameter.
Since the pairwise similarity among the code representation vectors and description representation vectors still cannot comprehensively present the distribution condition of them in the whole embedding space, we involve a matrix $\tilde{S}\tilde{S}^T$ to describe a high order neighborhood similarity that two vectors with high similarity should also have the close similarity to other vectors. 
Finally, we utilize a weighted equation to combine both of these two matrices as follows:
\begin{equation}
    S = (1-\eta)\tilde{S} + \eta \frac{\tilde{S}\tilde{S}^T}{m},
\end{equation}
\noindent where $\eta$ is a hyper-parameter and $m$ is the batch size which is utilized to normalize the second term in the equation. Since we hope the binary hash codes of the source code and its corresponding description to be the same, we replace the diagonal elements in the similarity matrix with one. The final high order similarity matrix is:
\begin{equation}
S_{F_{ij}}=
\begin{cases}
1,& i = j \\
S_{ij}, & \text{otherwise}
\end{cases}
\end{equation}

\noindent{\textbf{Binary Hash Code Training}}
We propose to replace the output layer of the original code search model with three FC layers with $\rm tanh(\cdot)$ activate function. We define the trained binary hash code for code and description as $B_C=\{b_{c}^{(1)},...,b_{c}^{(n)}\}$ and $B_D=\{b_{d}^{(1)},...,b_{d}^{(n)}\}$, respectively. To ensure that the relative distribution of binary hash codes is similar to the distribution of representation vectors in the original embedding space, the following equation is utilized as the loss function of the deep hashing module:
\begin{equation}
\begin{aligned}
\mathscr{L}(\theta) &   = \min_{B_C, B_D} \Vert \min (\mu S_F, 1) - \frac{B_C B_D^T}{d} \Vert_F^2 \\
  &   + \lambda_1 \Vert \min (\mu S_F, 1) - \frac{B_C B_C^T}{d} \Vert_F^2 \\
  &   + \lambda_2 \Vert \min (\mu S_F, 1) - \frac{B_D B_D^T}{d} \Vert_F^2, \\
 &  s.t. \ B_C, B_D \in \{-1,+1\}^{m \times d}, \\
\end{aligned}
\end{equation}
\noindent where $\theta$ are model parameters, $\mu$ is the weighted parameters to adjust the similarity score between different pairs of code and description, $\lambda_1$, $\lambda_2$ are the trade-off parameters to weight different terms in the loss function, and $d$ is the dimension of the binary hash code generated by this deep hashing module. These three terms in the loss function are adopted to restrict the similarity among binary hash codes of the source codes, the similarity among binary hash codes of the descriptions, and the similarity between the binary hash codes of source code and description, respectively. 

Note that we adopt $B_C B_D^T/d$ to replace ${\rm cos}(B_C, B_D)$ because ${\rm cos}(B_C, B_D)$ only measures the angle between two vectors but neglects the length of the vectors, which makes ${\rm cos}(B_C, B_D)$ can still be a very large value even the value of every hash bits is close to zero. Unlike ${\rm cos}(B_C, B_D)$, $B_C B_D^T/d$ can only achieve a high value when every bit of the binary hash code is 1 or -1 since the value of $B_C B_D^T/d$ will be close to zero if the value of every hash bits is close to zero.

Since it is impractical to impose on the output of neural network to be discrete values like 1 and -1, we adopt the following equation to convert the output of deep hashing module to be strict binary hash code:
\begin{equation}
B = {\rm sgn}(H) \in \{-1, +1\}^{m \times d},
\end{equation}
\noindent where $H$ is the output of the last hidden layer without the activation function in the deep hashing module and $\rm sgn(\cdot)$ is the sign function and the output of this function is 1 if the input is positive and the output is -1 otherwise. 

However, the gradient of the sign function will be zero in  backward propagation which will induce the vanishing gradients problem and affect model convergence. To address this problem, we follow the previous research~\citep{CaoLWY17, HuNL19} and adopt a scaling function:
\begin{equation}
B = {\rm tanh}(\alpha H) \in \{-1, +1\}^{m \times d},
\label{eq:restriction}
\end{equation}
\noindent where $\alpha$ is the parameter which is increased during the training. The function of ${\rm tanh}(\alpha H)$ is an approximate equation of ${\rm sgn}(H)$ when $\alpha$ is large enough. Therefore, the output of Eq.~\ref{eq:restriction} will finally be converged to 1 or -1 with the increasing of $\alpha$ during the training and the above problem is addressed.

\saveSpaceText
\subsection{Online Stage}
\saveSpaceText
\textbf{Recall and Re-rank Mechanism} The incoming query from users will be fed into the description category prediction module to calculate the normalized probability distribution of categories at first. Then the number of code candidates $R_i$ for each category $i$ will be determined according to this probability distribution. The Hamming distance between the hash code of the given query and all the code inside the database will be calculated. Then code candidates will be sorted by Hamming distance in ascending order and the top $R_i$ code candidates in each category $i$ will be recalled. In the re-rank step, the original representation vectors of these recalled code candidates will be retrieved and utilized for the cosine similarity calculation. Finally, code snippets will be returned to users in descending order of cosine similarity.

\noindent \textbf{Description Category Prediction Module} The description category prediction module aims to predict the category of source code that meets user's requirement according to the given natural language description. The model adopted for category prediction is the same as the original code search model, except that the output layer is replaced with a one-hot category prediction layer and the cross-entropy function is adopted as the loss function of the model. 

Since the accuracy of the description category prediction module is not perfect, we use the probability distribution of each category instead of the category with the highest predicted probability as the recall strategy for code search. We define the total recall number of source code as $N$, the normalized predicted probability for each code category as $P=\{p_1, ..., p_k\}$, where $k$ is the number of categories. The recall number of source code in each category is:
\begin{equation}
R_i = \min (\lfloor p_i \cdot (N-k) \rfloor, 1), \ i \in 1, ..., k,
\end{equation}
\noindent where $R_i$ is the recall number of source code in category $i$. To ensure that the proposed approach can recall at least one source code from each category, we set the minimum recall number for a single category to 1. 
\saveSpaceText
\section{Experiments}
\label{sec:experiment}
\saveSpaceText

\subsection{Dataset}
\label{sec:dataset}
We use two datasets (Python and Java) provided by CodeBERT~\citep{FengGTDFGS0LJZ20} to evaluate the performance of \tool. CodeBERT selects the data from the CodeSearchNet~\cite{abs-1909-09436} dataset and creates both  positive and negative examples of <description, code> pairs. Since all the baselines in our experiments are bi-encoder models, we do not need to predict the relevance score for the mismatched pairs so we remove all the negative examples from the dataset. Finally we get 412,178 <description, code> pairs as the training set, 23,107 <description, code> pairs as the validation set, and 22,176 <description, code> pairs as the test set in the Python dataset. We get 454,451 <description, code> pairs as the training set, 15,328 <description, code> pairs as the validation set, and 26,909 <description, code> pairs as the test set in the Java dataset.

\subsection{Experimental Setup}
In the code classification module, we set the number of cluster to 10. In the deep hashing module, we add three fully connected (FC) layer in all the baselines, the hidden size of each FC layer is the same as the dimension of the original representation vectors. Specifically, the hidden size of FC layer for CodeBERTa, CodeBERT, GraphCodeBERT is 768. The hidden size of FC layer for UNIF is 512 and for RNN is 2048. The size of the output binary hash code for all the baselines is 128. The hyper parameters $\beta, \eta, \mu, \lambda_1, \lambda_2$ are 0.6, 0.4, 1.5, 0.1, 0.1, respectively. The parameter $\alpha$ is the epoch number and will be linear increased during the training. In the query category prediction module,  a cross-entropy function is adopted as the loss function and the total recall number is 100. 

The learning rate for CodeBERTa, CodeBERT and GraphCodeBERT is 1e-5 and the learning rate for UNIF, RNN is 1.34e-4. All the models are trained via the AdamW algorithm~\citep{KingmaB14}.

We train our models on a server with four 4x Tesla V100 w/NVLink and 32GB memory. Each module based on CodeBERT, GraphCodeBERT and CodeBERTa are trained with 10 epochs and Each module based on RNN and UNIF are trained with 50 epochs. The early stopping strategy is adopted to avoid overfitting for all the baselines. The time efficiency experiment is conducted on the server with Intel Xeon E5-2698v4 2.2Ghz 20-core. The programming for evaluation is written in C++ and the program is allowed to use single thread of CPU.

\saveSpaceText
\subsection{Baselines}
\saveSpaceText
We apply \tool on several state-of-the-art and representative baseline models. UNIF~\citep{CambroneroLKS019} regards the code as the sequence of tokens and embeds the sequence of code tokens and description tokens into representation vectors via full connected layer with attention mechanism, respectively. 
RNN baseline adopts a two-layer bi-directional LSTM~\citep{ChoMBB14} to encode the input sequences. CodeBERTa~\footnote{\url{https://huggingface.co/huggingface/CodeBERTa-small-v1}} is a 6-layer, Transformer-based model trained on the CodeSearchNet dataset. CodeBERT~\citep{FengGTDFGS0LJZ20} is a pre-trained model based on Transformer with 12 layers. 
Similar to CodeBERT, GraphCodeBERT~\citep{GuoRLFT0ZDSFTDC21} is a pre-trained Transformer-based model pre-trained with not only tokens information but also dataflow of the code snippets. As we introduced, the inference efficiency of cross-encoder based models like CodeBERT is quite low and the purpose of our approach is to improve the calculation efficiency between the representation vectors of code and queries. Here we slightly change the model structure of CodeBERTa, CodeBERT, and GraphCodeBERT. Rather than concatenating code and query together and inputting them into a single encoder to predict the relevance score of the pair, we adopt the bi-encoder architecture for the baselines, which utilize the independent encoder to encoding the code and queries into representation vectors, respectively. Also, cosine similarity between the given representation vector pairs is adopted as the training loss function to replace the cross entropy function of the output relevance score.

\saveSpaceText
\subsection{Evaluation Metric}
\saveSpaceText
$SuccessRate@k$ is widely used by many previous  studies~\cite{haldar2020multi,shuai2020improving,fang2021self,heyman2020neural}. The metric is calculated as follows:
\begin{equation}
\small
    SuccessRate@k = \frac{1}{|Q|}\sum^Q_{q=1}\delta(FRank_q \leq k),
\end{equation}
\noindent where $Q$ denotes the query set and $FRank_q$ is the rank of the correct answer for query $q$. If the correct result is within the top $k$ returning results, $\delta(FRank_q \leq k)$ returns 1, otherwise it returns 0. A higher $R@k$ indicates better  performance.

\saveSpaceText
\subsection{Experimental Results}
\saveSpaceText
In this section, we present the experimental results and evaluate the performance of \tool from the aspects of retrieval efficiency, overall retrieval performance, and the effectiveness of the internal classification module.

\saveSpaceText
\subsubsection{RQ1: How much faster is \tool than the original code search models?}
\saveSpaceText

\begin{table}
\small
\centering
\begin{tabular}{lll}
\toprule
& \textbf{Python} & \textbf{Java}\\ \midrule 
& \multicolumn{2}{c}{\emph{Total Time}} \\ \cmidrule(r){2-3} 
CodeBERT & 572.97s & 247.78s \\
\tool & 33.87s ($\downarrow$94.09\%) & 15.78s ($\downarrow$93.51\%) \\ \midrule
& \multicolumn{2}{c}{ \emph{(1) Vector Similarity Calculation}} \\ \cmidrule(r){2-3} 
CodeBERT & 531.95s & 234.08s \\
\tool & 14.43s ($\downarrow$97.29\%) & 7.25s ($\downarrow$96.90\%) \\ \midrule
& \multicolumn{2}{c}{ \emph{(2) Array Sorting}} \\ \cmidrule(r){2-3} 
CodeBERT & 41.02s & 13.70s \\
\tool & 19.44s ($\downarrow$53.61\%) & 8.53s ($\downarrow$37.74\%) \\
\bottomrule
\end{tabular}
\saveSpaceFig
\caption{Time Efficiency of \tool.}
\saveSpaceFig
\label{tab:efficiency}
\end{table}

\begin{table*}[ht]
\footnotesize
\setlength\tabcolsep{2pt}
\centering
\begin{tabular}{lllllll}
\toprule
\multirow{2}{*}{\textbf{Model}} & \multicolumn{3}{c}{\textbf{Python}} & \multicolumn{3}{c}{\textbf{Java}} \\
\cmidrule(lr){2-4} \cmidrule(lr){5-7}
& \textbf{R@1} & \textbf{R@5} & \textbf{R@10} & \textbf{R@1} & \textbf{R@5} & \textbf{R@10}\\
\midrule
$\rm UNIF$ & 0.071 & 0.173 &0.236 & 0.084 & 0.193 & 0.254\\ 
\hdashline
$\rm \tool_{UNIF}$ & \textbf{0.072 ($\uparrow$1.4\%)} &  \textbf{0.177 ($\uparrow$2.3\%)} &  \textbf{0.241 ($\uparrow$2.1\%)} &  \textbf{0.086 ($\uparrow$2.4\%)} & \textbf{0.198 ($\uparrow$2.6\%)} & \textbf{0.264 ($\uparrow$3.9\%)} \\
$\rm \ \ -w/o \ classification $ & 0.071 (0.0\%) &  0.174 ($\uparrow$0.6\%) &  0.236 (0.0\%) & 0.085 ($\uparrow$1.2\%) & 0.193 (0.0\%) & 0.254 (0.0\%) \\
$\rm \ \ -one \ classification$ & 0.069 ($\downarrow$2.8\%) & 0.163 ($\downarrow$5.8\%) & 0.216 ($\downarrow$8.5\%)  & 0.083 ($\downarrow$1.2\%)  & 0.183 ($\downarrow$5.2\%) & 0.236 ($\downarrow$7.1\%)  \\
\hdashline
$\rm \ \  -ideal \ classification$ & 0.077 ($\uparrow$6.9\%) & 0.202 ($\uparrow$16.8\%) & 0.277 ($\uparrow$17.4\%) & 0.093 ($\uparrow$10.7\%) & 0.222 ($\uparrow$15.0\%) & 0.296  ($\uparrow$16.5\%) \\

\midrule
$\rm RNN$ & 0.111 & 0.253 & 0.333 & 0.073 & 0.184 & 0.250\\ \hdashline
$\rm \tool_{RNN}$ & \textbf{0.112 ($\uparrow$0.9\%)} &  \textbf{0.259 ($\uparrow$2.4\%)} &  \textbf{0.343 ($\uparrow$5.0\%)} &  \textbf{0.076 ($\uparrow$4.1\%)} & \textbf{0.194 ($\uparrow$5.4\%)} & \textbf{0.265 ($\uparrow$6.0\%)} \\
$\rm \ \ -w/o \ classification$ & \textbf{0.112 ($\uparrow$0.9\%)} & 0.254 ($\uparrow$0.4\%) & 0.335 ($\uparrow$0.6\%) & 0.073 (0.0\%) & 0.186 ($\uparrow$1.1\%) & 0.253 ($\uparrow$1.2\%)  \\
$\rm \ \ -one \ classification$ & \textbf{0.112 ($\uparrow$0.9\%)} & 0.243 ($\downarrow$4.0\%) & 0.311 ($\downarrow$6.6\%) & 0.075 ($\uparrow$2.7\%) & 0.182 ($\downarrow$1.1\%) & 0.240 ($\downarrow$4.0\%) \\
\hdashline
$\rm \ \ -ideal \ classification$ & 0.123 ($\uparrow$10.8\%) & 0.289 ($\uparrow$14.2\%) & 0.385  ($\uparrow$15.6\%) & 0.084 ($\uparrow$15.1\%) & 0.221 ($\uparrow$20.1\%) & 0.302 ($\uparrow$20.8\%) \\

\midrule
$\rm CodeBERTa$ & 0.124 & 0.250 & 0.314 & 0.089 & 0.203 & 0.264\\ \hdashline
$\rm \tool_{CodeBERTa}$ & \textbf{0.123 ($\downarrow$0.8\%)} &  \textbf{0.247 ($\downarrow$1.2\%)} &  \textbf{0.309 ($\downarrow$1.6\%)} &  \textbf{0.090 ($\uparrow$1.1\%)} & \textbf{0.210 ($\uparrow$3.4\%)} & \textbf{0.272 (($\uparrow$3.0\%)} \\
$\rm \ \ -w/o \ classification$ & 0.122 ($\downarrow$1.6\%) & 0.242 ($\downarrow$3.2\%) & 0.302 ($\downarrow$3.8\%)& 0.089 (0.0\%) & 0.201 ($\downarrow$1.0\%) & 0.258 ($\downarrow$2.3\%) \\
$\rm \ \ -one \ classification$ & 0.116 ($\downarrow$6.5\%) & 0.221 ($\downarrow$11.6\%) & 0.271 ($\downarrow$13.7\%)  & 0.085  ($\downarrow$4.5\%)  & 0.189 ($\downarrow$6.9\%) & 0.238 ($\downarrow$9.8\%)  \\
\hdashline
$\rm \ \ -ideal \ classification$ & 0.135 ($\uparrow$8.9\%) & 0.276 ($\uparrow$10.4\%) & 0.346  ($\uparrow$10.2\%) & 0.100 ($\uparrow$12.4\%) & 0.235 ($\uparrow$15.8\%) & 0.305 ($\uparrow$15.5\%) \\

\midrule
$\rm CodeBERT$ & 0.451 & 0.683 & 0.759 & 0.319 & 0.537 & 0.608\\ \hdashline
$\rm \tool_{CodeBERT}$ & \textbf{0.451 (0.0\%)} &  \textbf{0.679 ($\downarrow$0.6\%)} &  \textbf{0.750 ($\downarrow$1.2\%)} &  \textbf{0.318 ($\downarrow$0.3\%)} & \textbf{0.533 ($\downarrow$0.7\%)} & \textbf{0.602 ($\downarrow$1.0\%)} \\
$\rm \ \ -w/o \ classification$ & 0.449 ($\downarrow$0.4\%) & 0.673 ($\downarrow$1.5\%) & 0.742 ($\downarrow$2.2\%) & 0.316 ($\downarrow$0.9\%) & 0.527 ($\downarrow$1.9\%) & 0.593 ($\downarrow$2.5\%)\\
$\rm \ \ -one \ classification$ & 0.425 ($\downarrow$5.8\%) & 0.613 ($\downarrow$10.2\%) & 0.665 ($\downarrow$12.4\%)  & 0.304 ($\downarrow$4.7\%)  & 0.483 ($\downarrow$10.1\%) & 0.532 ($\downarrow$12.5\%)  \\
\hdashline
$\rm \ \ -ideal \ classification$ & 0.460 ($\uparrow$2.0\%) & 0.703 ($\uparrow$2.9\%) & 0.775  ($\uparrow$2.1\%) & 0.329 ($\uparrow$3.1\%) & 0.555 ($\uparrow$3.4\%) & 0.627  ($\uparrow$3.1\%) \\

\midrule
$\rm GraphCodeBERT$ & 0.485 & 0.726 & 0.792 & 0.353 & 0.571 & 0.640\\ \hdashline
$\rm \tool_{GraphCodeBERT}$ & \textbf{0.483 ($\downarrow$0.4\%)} &  \textbf{0.719 ($\downarrow$1.0\%)} &  \textbf{0.782 ($\downarrow$1.3\%)} &  \textbf{0.350 ($\downarrow$0.8\%)} & \textbf{0.561 ($\downarrow$1.8\%)} & \textbf{0.625 ($\downarrow$2.3\%)} \\
$\rm \ \ -w/o \ classification$ & 0.481 ($\downarrow$0.8\%) & 0.713 ($\downarrow$1.8\%) & 0.774 ($\downarrow$2.3\%) & 0.347 ($\downarrow$1.7\%) & 0.553 ($\downarrow$3.2\%) & 0.616 ($\downarrow$3.7\%) \\
$\rm \ -one \ classification$ & 0.459 ($\downarrow$5.4\%) & 0.653 ($\downarrow$10.1\%) & 0.698 ($\downarrow$11.9\%)  & 0.329 ($\downarrow$7.8\%)  & 0.505 ($\downarrow$11.6\%) & 0.551 ($\downarrow$13.9\%)  \\
\hdashline
$\rm \ \ -ideal\ classification$ & 0.494 ($\uparrow$1.9\%) & 0.741 ($\uparrow$2.1\%) & 0.803  ($\uparrow$1.4\%) &  0.361 ($\uparrow$2.3\%) & 0.585 ($\uparrow$2.5\%) & 0.649 ($\uparrow$1.4\%) \\
\bottomrule
\end{tabular}
\saveSpaceFig
\caption{Results of code search performance comparison. The best results among the three \tool variants are highlighted in \textbf{bold} font.}
\saveSpaceFig
\label{tab:overall}
\end{table*}

Table~\ref{tab:efficiency} illustrates the results of efficiency comparison between the original code search models and \tool. Once the representation vectors of code and description are stored in the memory, the retrieval efficiency mainly depends on the dimension of representation vectors rather than the complexity of the original retrieval model. Therefore, we select CodeBERT as the baseline model to illustrate efficiency comparison. Since code search process in both approaches contains vector similarity calculation and array sorting, we split the retrieval process into these two steps to calculate the time cost. 

In the vector similarity calculation step, \tool reduces 97.29\% and 96.90\% of time cost in the dataset of Python and Java respectively, which demonstrates that the utilization of binary hash code can effectively reduce vector similarity calculation cost in the code retrieval process.

In the array sorting step, \tool reduces 53.61\% and 37.74\% of time cost in the dataset of Python and Java, respectively. The classification module makes the main contribution on the improvement of sorting efficiency. The sorting algorithm applied in both original code search model and \tool is quick sort, whose time complexity is $O(n{\rm log}n)$. Classification module divides a large code dataset into several small code datasets, reducing the average time complexity of sorting to $O(n{\rm log}\frac{n}{m})$. The reason why the improvement of sorting in the Java dataset is not so significant as in the Python dataset is that the size of Java dataset is much smaller than the size of Python dataset. However, the combination of the algorithm of divide and conquer and max-heap, rather than quick sort, is widely applied in the big data sorting, which can greatly shrink the retrieval efficiency gap between these two approaches. Therefore, the improvement of efficiency in the sorting process will not be as large as what shown in Table~\ref{tab:efficiency}.

In the overall code retrieval process, the cost time is reduced by 94.09\% and 93.51\% in the dataset of Python and Java, respectively. Since the vector similarity calculation takes most of cost time in the code retrieval process, \tool still can reduce at least 90\% of cost time, which demonstrates the effectiveness on the efficiency improvement in the code search task. 

\saveSpaceText
\subsubsection{RQ2: How does \tool affect the accuracy of the original models? }
\label{sec:RQ2}
\saveSpaceText
Table~\ref{tab:overall} illustrates the retrieval performance comparison  between the original code search models and \tool. We have noticed that the performance of the conventional approaches like BM25~\citep{RobertsonZ09} is not good enough. For example, we set the token length for both code and queries as 50, which is the same as the setting in CodeBERT, and apply BM25 to recall top 100 code candidates for the re-rank step on the Python dataset. BM25 can only retain 99.3\%, 95.6\% and 92.4\% retrieval accuracy of CodeBERT in terms of $R@1$, $R@5$ and $R@10$ on the Python dataset. Here we only compare the performance of our approach with the original code search models since the purpose of our approach is to preserve the performance of the original code search models.
As can be observed, \tool can retain at least 99.5\%, 99.0\% and 98.4\% retrieval accuracy of most original code search models in terms of $R@1$, $R@5$ and $R@10$
on the Python dataset. \tool can also retain 99.2\%, 98.2\% and 97.7\% of the retrieval accuracy as all original code search baselines in terms of $R@1$, $R@5$ and $R@10$ on the Java dataset, respectively.
We can find that \tool can retain more than 97.7\% of performance in all metrics. $R@1$ is the most important and useful metric among these metrics since most users hope that the first returned answer is the correct answer during the search. \tool can retain at least 99.2\% of performance on $R@1$ in both datasets, which demonstrates that \tool can retain almost the same performance as the original code search model.

It is interesting that \tool presents a relatively better performance when the performance of the original code retrieval models is worse. $\rm \tool_{CodeBERTa}$ even outperforms the original baseline model in Java dataset. $\rm \tool_{RNN}$ and $\rm \tool_{UNIF}$ outperform the original model in both Python and Java datasets. The integration of deep learning and code classification with recall make the contribution on this result. The worse performance indicates more misalignment between the code representation vectors and description representation vectors. Since the code classification and deep hashing will filter out most of irrelevant codes in the recall stage, some irrelevant code representation vectors but has high cosine similarity with the target description representation vectors are filtered, which leads the improvement on the final retrieval performance.

\saveSpaceText
\subsubsection{RQ3: Can the classification module help improve performance?}
\label{sec:RQ3}
\saveSpaceText

\begin{table}
\small
\centering
\begin{tabular}{lll}
\toprule
 \multirow{2}{*}{\textbf{Model}} &  \textbf{Python} &  \textbf{Java}\\ 
& \textbf{Acc.} & \textbf{Acc.} \\
\midrule 
$\rm \tool_{UNIF}$ & 0.558 & 0.545\\
$\rm \tool_{RNN}$ & 0.610 & 0.535  \\
$\rm \tool_{CodeBERTa}$ & 0.591  & 0.571  \\
$\rm \tool_{CodeBERT}$ & 0.694  & 0.657 \\
$\rm \tool_{GraphCodeBERT}$ & 0.713  & 0.653 \\
\bottomrule
\end{tabular}
\saveSpaceFig
\caption{Classification accuracy of the code classification module in each model.}
\saveSpaceFig
\label{tab:classification_accuray}
\end{table}

Table~\ref{tab:overall} illustrates the performance comparison between the \tool variants which adopt different recall strategies with query category prediction results. 
$\rm \tool_{w/o \ classification}$ represents \tool without code classification and description prediction module. $\rm \tool_{one \ classification}$ represents the \tool variant that recalls $N-k+1$  candidates in the code category with highest prediction probability and  one in each of the rest categories. $\rm \tool_{ideal \ classification}$ is an ideal classification situation we set. Assuming the correct description category is known, $N-k+1$ candidates are recalled in the correct category and one candidate is recalled in each of the rest categories. Note that the display of $\rm \tool_{ideal \ classification}$ is only to explore the upper threshold of performance improvement of the category prediction module and will not be counted as a variant of $\rm \tool$ we compare.

By comparing the performance between $\rm \tool_{ideal \ classification}$  and $\rm \tool_{w/o \ classification}$, we can find that correct classification can significantly improve the retrieval performance. With the ideal  category labels, \tool can even outperform all baseline models. As mentioned in \Sec~\ref{sec:RQ2}, code classification can mitigate the problem of vector pairs misalignment via filtering out wrong candidates whose representation vectors has high cosine similarity with the target representation vectors in the recall stage. The more serious the misalignment problem, the more effective the code classification. That is the reason why the improvement of \tool with ground-truth labels on UNIF, RNN, and CodeBERTa is more significant than the improvement of it on CodeBERT and GraphCodeBERT since the retrieval accuracy of former models is much lower than the latter ones. Similar conclusions can also be drawn at the aspect of binary hash code distribution via the comparison between \tool and $\rm \tool_{ideal \ classification}$ since \tool utilizes the distribution of the original representation vectors as the guidance for model training. Therefore, the distribution of binary hash codes will be similar to the distribution of original representation vectors. 

Since we have explored the theoretical upper limit of the effectiveness of  code classification for code retrieval, the effectiveness of  code classification for code retrieval in the real application will be validated. By comparing the experimental results between $\rm \tool_{w/o \ classification}$ and $\rm \tool_{one \ classification}$, we can find that the performance of \tool with predicted labels is even worse than the performance of \tool without code classification module. The reason is that the accuracy of description category prediction is far from the satisfactory. Table~\ref{tab:classification_accuray} illustrates the accuracy of description category prediction module in all baseline models. We regard the category with the highest probability as the predicted category from the description category prediction module and check whether the module could give a correct prediction. It can be seen that the classification accuracy is not very high (less than 75\%). By observing the experimental results of \tool in GraphCodeBERT on the Java dataset, we can also find that low accuracy greatly affect the performance of  $\rm \tool_{one classification}$, which makes 7.8\%, 11.6\%, and 13.9\% performance drop in terms of $R@1$, $R@5$, and $R@10$, respectively.

Fortunately, although the description category prediction module cannot accurately tell the exact category which this description belongs to, the module still can give a relative high predicted probability on the correct category. By comparing the experimental results among all the variants of \tool, we can find the performance is increased significantly once the recall strategy is replaced to that the number of code candidates for each category is determined by the normalized predication probability. \tool with new recall strategy almost achieve the best performance in all metrics on all baseline models. Even on RNN in the Python dataset, \tool still achieve the same performance as \tool without classification under $R@1$ and achieve similar performance in other metrics. Above experimental results have demonstrated the effectiveness of the adoption of code classification in code search.

\saveSpaceText
\section{Conclusion}
\label{sec:conclusion}
\saveSpaceText
\vspace{-3pt}

To accelerate code search, we present \tool, a general method that incorporates deep hashing techniques and code classification. We leverage the two-staged recall and re-rank paradigm in information retrieval field and apply deep hashing techniques for fast recall. Furthermore, we propose to utilize a code classification module to retrieve better quality code snippets. Experiments on five code search models show that compared with the original code search models, \tool can greatly improve the retrieval efficiency meanwhile preserve almost the same performance.
\section{Acknowledgement}
\label{sec:acknowledgement}

Wenchao Gu’s and Michael R. Lyu’s work described in this paper was in part supported by the Research Grants Council of the Hong Kong Special Administrative Region, China (CUHK 14210920 of the General Research Fund). 

\bibliography{custom}

\begin{thebibliography}{30}
\expandafter\ifx\csname natexlab\endcsname\relax\def\natexlab#1{#1}\fi

\bibitem[{Brandt et~al.(2009)Brandt, Guo, Lewenstein, Dontcheva, and
  Klemmer}]{BrandtGLDK09}
Joel Brandt, Philip~J. Guo, Joel Lewenstein, Mira Dontcheva, and Scott~R.
  Klemmer. 2009.
\newblock \href {https://doi.org/10.1145/1518701.1518944} {Two studies of
  opportunistic programming: interleaving web foraging, learning, and writing
  code}.
\newblock In \emph{Proceedings of the 27th International Conference on Human
  Factors in Computing Systems, {CHI} 2009, Boston, MA, USA, April 4-9, 2009},
  pages 1589--1598. {ACM}.

\bibitem[{Cambronero et~al.(2019)Cambronero, Li, Kim, Sen, and
  Chandra}]{CambroneroLKS019}
Jos{\'{e}} Cambronero, Hongyu Li, Seohyun Kim, Koushik Sen, and Satish Chandra.
  2019.
\newblock \href {https://doi.org/10.1145/3338906.3340458} {When deep learning
  met code search}.
\newblock In \emph{Proceedings of the {ACM} Joint Meeting on European Software
  Engineering Conference and Symposium on the Foundations of Software
  Engineering, {ESEC/SIGSOFT} {FSE} 2019, Tallinn, Estonia, August 26-30,
  2019}, pages 964--974. {ACM}.

\bibitem[{Cao et~al.(2017)Cao, Long, Wang, and Yu}]{CaoLWY17}
Zhangjie Cao, Mingsheng Long, Jianmin Wang, and Philip~S. Yu. 2017.
\newblock \href {https://doi.org/10.1109/ICCV.2017.598} {Hashnet: Deep learning
  to hash by continuation}.
\newblock In \emph{{IEEE} International Conference on Computer Vision, {ICCV}
  2017, Venice, Italy, October 22-29, 2017}, pages 5609--5618. {IEEE} Computer
  Society.

\bibitem[{Cho et~al.(2014)Cho, van Merrienboer, Bahdanau, and
  Bengio}]{ChoMBB14}
Kyunghyun Cho, Bart van Merrienboer, Dzmitry Bahdanau, and Yoshua Bengio. 2014.
\newblock \href {https://doi.org/10.3115/v1/W14-4012} {On the properties of
  neural machine translation: Encoder-decoder approaches}.
\newblock In \emph{Proceedings of SSST@EMNLP 2014, Eighth Workshop on Syntax,
  Semantics and Structure in Statistical Translation, Doha, Qatar, 25 October
  2014}, pages 103--111. Association for Computational Linguistics.

\bibitem[{Ding et~al.(2014)Ding, Guo, and Zhou}]{DingGZ14}
Guiguang Ding, Yuchen Guo, and Jile Zhou. 2014.
\newblock \href {https://doi.org/10.1109/CVPR.2014.267} {Collective matrix
  factorization hashing for multimodal data}.
\newblock In \emph{2014 {IEEE} Conference on Computer Vision and Pattern
  Recognition, {CVPR} 2014, Columbus, OH, USA, June 23-28, 2014}, pages
  2083--2090. {IEEE} Computer Society.

\bibitem[{Du et~al.(2021)Du, Shi, Wang, Shi, Han, and Zhang}]{du2021single}
Lun Du, Xiaozhou Shi, Yanlin Wang, Ensheng Shi, Shi Han, and Dongmei Zhang.
  2021.
\newblock \href {https://doi.org/10.1145/3459637.3482127} {Is a single model
  enough? mucos: {A} multi-model ensemble learning approach for semantic code
  search}.
\newblock In \emph{{CIKM} '21: The 30th {ACM} International Conference on
  Information and Knowledge Management, Virtual Event, Queensland, Australia,
  November 1 - 5, 2021}, pages 2994--2998. {ACM}.

\bibitem[{Fang et~al.(2021)Fang, Tan, Zhang, and Liu}]{fang2021self}
Sen Fang, Youshuai Tan, Tao Zhang, and Yepang Liu. 2021.
\newblock \href {https://doi.org/10.1016/j.infsof.2021.106542} {Self-attention
  networks for code search}.
\newblock \emph{Inf. Softw. Technol.}, 134:106542.

\bibitem[{Feng et~al.(2020)Feng, Guo, Tang, Duan, Feng, Gong, Shou, Qin, Liu,
  Jiang, and Zhou}]{FengGTDFGS0LJZ20}
Zhangyin Feng, Daya Guo, Duyu Tang, Nan Duan, Xiaocheng Feng, Ming Gong, Linjun
  Shou, Bing Qin, Ting Liu, Daxin Jiang, and Ming Zhou. 2020.
\newblock \href {https://doi.org/10.18653/v1/2020.findings-emnlp.139}
  {Codebert: {A} pre-trained model for programming and natural languages}.
\newblock In \emph{Findings of the Association for Computational Linguistics:
  {EMNLP} 2020, Online Event, 16-20 November 2020}, volume {EMNLP} 2020 of
  \emph{Findings of {ACL}}, pages 1536--1547. Association for Computational
  Linguistics.

\bibitem[{Gu et~al.(2021)Gu, Li, Gao, Wang, Zhang, Xu, and Lyu}]{GuLGWZXL21}
Wenchao Gu, Zongjie Li, Cuiyun Gao, Chaozheng Wang, Hongyu Zhang, Zenglin Xu,
  and Michael~R. Lyu. 2021.
\newblock \href {https://doi.org/10.1016/j.neunet.2021.04.019} {Cradle: Deep
  code retrieval based on semantic dependency learning}.
\newblock \emph{Neural Networks}, 141:385--394.

\bibitem[{Gu et~al.(2018)Gu, Zhang, and Kim}]{GuZ018}
Xiaodong Gu, Hongyu Zhang, and Sunghun Kim. 2018.
\newblock \href {https://doi.org/10.1145/3180155.3180167} {Deep code search}.
\newblock In \emph{Proceedings of the 40th International Conference on Software
  Engineering, {ICSE} 2018, Gothenburg, Sweden, May 27 - June 03, 2018}, pages
  933--944. {ACM}.

\bibitem[{Guo et~al.(2021)Guo, Ren, Lu, Feng, Tang, Liu, Zhou, Duan,
  Svyatkovskiy, Fu, Tufano, Deng, Clement, Drain, Sundaresan, Yin, Jiang, and
  Zhou}]{GuoRLFT0ZDSFTDC21}
Daya Guo, Shuo Ren, Shuai Lu, Zhangyin Feng, Duyu Tang, Shujie Liu, Long Zhou,
  Nan Duan, Alexey Svyatkovskiy, Shengyu Fu, Michele Tufano, Shao~Kun Deng,
  Colin~B. Clement, Dawn Drain, Neel Sundaresan, Jian Yin, Daxin Jiang, and
  Ming Zhou. 2021.
\newblock \href {https://openreview.net/forum?id=jLoC4ez43PZ} {Graphcodebert:
  Pre-training code representations with data flow}.
\newblock In \emph{9th International Conference on Learning Representations,
  {ICLR} 2021, Virtual Event, Austria, May 3-7, 2021}. OpenReview.net.

\bibitem[{Haldar et~al.(2020)Haldar, Wu, Xiong, and
  Hockenmaier}]{haldar2020multi}
Rajarshi Haldar, Lingfei Wu, Jinjun Xiong, and Julia Hockenmaier. 2020.
\newblock \href {https://doi.org/10.18653/v1/2020.acl-main.758} {A
  multi-perspective architecture for semantic code search}.
\newblock In \emph{Proceedings of the 58th Annual Meeting of the Association
  for Computational Linguistics, {ACL} 2020, Online, July 5-10, 2020}, pages
  8563--8568. Association for Computational Linguistics.

\bibitem[{He et~al.(2017)He, Xu, Lu, Yang, Shen, and Shen}]{0001XLYSS17}
Li~He, Xing Xu, Huimin Lu, Yang Yang, Fumin Shen, and Heng~Tao Shen. 2017.
\newblock \href {https://doi.org/10.1109/ICME.2017.8019549} {Unsupervised
  cross-modal retrieval through adversarial learning}.
\newblock In \emph{2017 {IEEE} International Conference on Multimedia and Expo,
  {ICME} 2017, Hong Kong, China, July 10-14, 2017}, pages 1153--1158. {IEEE}
  Computer Society.

\bibitem[{Heyman and Cutsem(2020)}]{heyman2020neural}
Geert Heyman and Tom~Van Cutsem. 2020.
\newblock \href {http://arxiv.org/abs/2008.12193} {Neural code search
  revisited: Enhancing code snippet retrieval through natural language intent}.
\newblock \emph{CoRR}, abs/2008.12193.

\bibitem[{Hu et~al.(2019)Hu, Nie, and Li}]{HuNL19}
Di~Hu, Feiping Nie, and Xuelong Li. 2019.
\newblock \href {https://doi.org/10.1109/TMM.2018.2866771} {Deep binary
  reconstruction for cross-modal hashing}.
\newblock \emph{{IEEE} Trans. Multim.}, 21(4):973--985.

\bibitem[{Husain et~al.(2019)Husain, Wu, Gazit, Allamanis, and
  Brockschmidt}]{abs-1909-09436}
Hamel Husain, Ho{-}Hsiang Wu, Tiferet Gazit, Miltiadis Allamanis, and Marc
  Brockschmidt. 2019.
\newblock \href {http://arxiv.org/abs/1909.09436} {Codesearchnet challenge:
  Evaluating the state of semantic code search}.
\newblock \emph{CoRR}, abs/1909.09436.

\bibitem[{Kingma and Ba(2015)}]{KingmaB14}
Diederik~P. Kingma and Jimmy Ba. 2015.
\newblock \href {http://arxiv.org/abs/1412.6980} {Adam: {A} method for
  stochastic optimization}.
\newblock In \emph{3rd International Conference on Learning Representations,
  {ICLR} 2015, San Diego, CA, USA, May 7-9, 2015, Conference Track
  Proceedings}.

\bibitem[{Luo et~al.(2020)Luo, Chen, Zhong, Zhang, Deng, Huang, and
  Hua}]{abs-2003-03369}
Xiao Luo, Chong Chen, Huasong Zhong, Hao Zhang, Minghua Deng, Jianqiang Huang,
  and Xiansheng Hua. 2020.
\newblock \href {http://arxiv.org/abs/2003.03369} {A survey on deep hashing
  methods}.
\newblock \emph{CoRR}, abs/2003.03369.

\bibitem[{Lv et~al.(2015)Lv, Zhang, Lou, Wang, Zhang, and Zhao}]{LvZLWZZ15}
Fei Lv, Hongyu Zhang, Jian{-}Guang Lou, Shaowei Wang, Dongmei Zhang, and
  Jianjun Zhao. 2015.
\newblock \href {https://doi.org/10.1109/ASE.2015.42} {Codehow: Effective code
  search based on {API} understanding and extended boolean model {(E)}}.
\newblock In \emph{30th {IEEE/ACM} International Conference on Automated
  Software Engineering, {ASE} 2015, Lincoln, NE, USA, November 9-13, 2015},
  pages 260--270. {IEEE} Computer Society.

\bibitem[{McMillan et~al.(2011)McMillan, Grechanik, Poshyvanyk, Xie, and
  Fu}]{McMillanGPXF11}
Collin McMillan, Mark Grechanik, Denys Poshyvanyk, Qing Xie, and Chen Fu. 2011.
\newblock \href {https://doi.org/10.1145/1985793.1985809} {Portfolio: finding
  relevant functions and their usage}.
\newblock In \emph{Proceedings of the 33rd International Conference on Software
  Engineering, {ICSE} 2011, Waikiki, Honolulu , HI, USA, May 21-28, 2011},
  pages 111--120. {ACM}.

\bibitem[{Robertson and Zaragoza(2009)}]{RobertsonZ09}
Stephen~E. Robertson and Hugo Zaragoza. 2009.
\newblock \href {https://doi.org/10.1561/1500000019} {The probabilistic
  relevance framework: {BM25} and beyond}.
\newblock \emph{Found. Trends Inf. Retr.}, 3(4):333--389.

\bibitem[{Sachdev et~al.(2018)Sachdev, Li, Luan, Kim, Sen, and
  Chandra}]{SachdevLLKS018}
Saksham Sachdev, Hongyu Li, Sifei Luan, Seohyun Kim, Koushik Sen, and Satish
  Chandra. 2018.
\newblock \href {https://doi.org/10.1145/3211346.3211353} {Retrieval on source
  code: a neural code search}.
\newblock In \emph{Proceedings of the 2nd {ACM} {SIGPLAN} International
  Workshop on Machine Learning and Programming Languages, MAPL@PLDI 2018,
  Philadelphia, PA, USA, June 18-22, 2018}, pages 31--41. {ACM}.

\bibitem[{Shuai et~al.(2020)Shuai, Xu, Liu, Yan, Xia, and
  Lei}]{shuai2020improving}
Jianhang Shuai, Ling Xu, Chao Liu, Meng Yan, Xin Xia, and Yan Lei. 2020.
\newblock \href {https://doi.org/10.1145/3387904.3389269} {Improving code
  search with co-attentive representation learning}.
\newblock In \emph{{ICPC} '20: 28th International Conference on Program
  Comprehension, Seoul, Republic of Korea, July 13-15, 2020}, pages 196--207.
  {ACM}.

\bibitem[{Su et~al.(2019)Su, Zhong, and Zhang}]{SuZZ19}
Shupeng Su, Zhisheng Zhong, and Chao Zhang. 2019.
\newblock \href {https://doi.org/10.1109/ICCV.2019.00312} {Deep joint-semantics
  reconstructing hashing for large-scale unsupervised cross-modal retrieval}.
\newblock In \emph{2019 {IEEE/CVF} International Conference on Computer Vision,
  {ICCV} 2019, Seoul, Korea (South), October 27 - November 2, 2019}, pages
  3027--3035. {IEEE}.

\bibitem[{Wang et~al.(2016)Wang, Liu, Kumar, and Chang}]{WangLKC16}
Jun Wang, Wei Liu, Sanjiv Kumar, and Shih{-}Fu Chang. 2016.
\newblock \href {https://doi.org/10.1109/JPROC.2015.2487976} {Learning to hash
  for indexing big data - {A} survey}.
\newblock \emph{Proc. {IEEE}}, 104(1):34--57.

\bibitem[{Wang et~al.(2014)Wang, Ooi, Yang, Zhang, and Zhuang}]{WangOYZZ14}
Wei Wang, Beng~Chin Ooi, Xiaoyan Yang, Dongxiang Zhang, and Yueting Zhuang.
  2014.
\newblock \href {https://doi.org/10.14778/2732296.2732301} {Effective
  multi-modal retrieval based on stacked auto-encoders}.
\newblock \emph{Proc. {VLDB} Endow.}, 7(8):649--660.

\bibitem[{Wu et~al.(2018)Wu, Lin, Han, Liu, Ding, Zhang, and Shen}]{WuLHLDZS18}
Gengshen Wu, Zijia Lin, Jungong Han, Li~Liu, Guiguang Ding, Baochang Zhang, and
  Jialie Shen. 2018.
\newblock \href {https://doi.org/10.24963/ijcai.2018/396} {Unsupervised deep
  hashing via binary latent factor models for large-scale cross-modal
  retrieval}.
\newblock In \emph{Proceedings of the Twenty-Seventh International Joint
  Conference on Artificial Intelligence, {IJCAI} 2018, July 13-19, 2018,
  Stockholm, Sweden}, pages 2854--2860. ijcai.org.

\bibitem[{Yao et~al.(2019)Yao, Peddamail, and Sun}]{YaoPS19}
Ziyu Yao, Jayavardhan~Reddy Peddamail, and Huan Sun. 2019.
\newblock \href {https://doi.org/10.1145/3308558.3313632} {Coacor: Code
  annotation for code retrieval with reinforcement learning}.
\newblock In \emph{The World Wide Web Conference, {WWW} 2019, San Francisco,
  CA, USA, May 13-17, 2019}, pages 2203--2214. {ACM}.

\bibitem[{Zhang et~al.(2018)Zhang, Peng, and Yuan}]{ZhangPY18}
Jian Zhang, Yuxin Peng, and Mingkuan Yuan. 2018.
\newblock \href
  {https://www.aaai.org/ocs/index.php/AAAI/AAAI18/paper/view/16746}
  {Unsupervised generative adversarial cross-modal hashing}.
\newblock In \emph{Proceedings of the Thirty-Second {AAAI} Conference on
  Artificial Intelligence, (AAAI-18), the 30th innovative Applications of
  Artificial Intelligence (IAAI-18), and the 8th {AAAI} Symposium on
  Educational Advances in Artificial Intelligence (EAAI-18), New Orleans,
  Louisiana, USA, February 2-7, 2018}, pages 539--546. {AAAI} Press.

\bibitem[{Zhou et~al.(2014)Zhou, Ding, and Guo}]{ZhouDG14}
Jile Zhou, Guiguang Ding, and Yuchen Guo. 2014.
\newblock \href {https://doi.org/10.1145/2600428.2609610} {Latent semantic
  sparse hashing for cross-modal similarity search}.
\newblock In \emph{The 37th International {ACM} {SIGIR} Conference on Research
  and Development in Information Retrieval, {SIGIR} '14, Gold Coast , QLD,
  Australia - July 06 - 11, 2014}, pages 415--424. {ACM}.

\end{thebibliography}
\bibliographystyle{acl_natbib}


\end{document}